\documentstyle [12pt,axodraw] {article}

\catcode`@=12
\begin{document}
\thispagestyle{empty}
\begin{flushright} UCRHEP-T247\\April 1999\
\end{flushright}
\vspace{0.5in}
\begin{center}
{\Large \bf Electron-Neutrino Majorana Mass\\
and Solar Neutrino Oscillations\\}
\vspace{1.3in}
{\bf Ernest Ma\\}
\vspace{0.3in}
{\sl Department of Physics\\}
{\sl University of California\\}
{\sl Riverside, California 92521\\}
\vspace{1.3in}
\end{center}
\begin{abstract}\ 
Assuming the Majorana masses of $\nu_e$ and the linear combination 
$c \nu_\tau - s \nu_\mu$ to be equal from their couplings to a heavy scalar 
triplet $\xi$, I show that their radiative splitting is given by 
$\Delta m^2/m_\nu^2 = (3 c^2/4 \pi^2) (G_F/\sqrt 2) m_\tau^2 
\ln (m_\xi^2/m_W^2)$.  This is applicable to the small-angle matter-enhanced 
oscillation solution of the solar neutrino deficit and restricts $m_\nu$ to 
be between 0.20 and 0.36 eV if $c^2 = 0.7$ and $m_\xi = 10^{14}$ GeV.
\end{abstract}
\newpage
\baselineskip 24pt

The existence of a small (and presumably Majorana) mass for each neutrino 
is now considered a near certainty, given the recent experimental evidence 
of neutrino oscillations in atmospheric \cite{1}, solar \cite{2}, and 
accelerator \cite{3} data.  However, since the above experiments only 
measure the differences of the squares of the neutrino masses, the absolute 
magnitude of each mass is undetermined.  In direct measurements of the 
electron-neutrino mass $m_{\nu_e}$, the best published upper limit at 
present is 4.35 eV \cite{4}.  In neutrinoless double beta decay, where 
only the effective mass
\begin{equation}
\langle m_{\nu_e} \rangle = | \Sigma ~U_{ei}^2 m_{\nu_i} |
\end{equation}
is measured, the best published upper limit at present is 0.46 eV \cite{5}. 
However, this has recently been reduced to 0.2 eV \cite{new}, and further 
significant improvement to much lower values is possible in the 
future \cite{6}.  If neutrino masses are hierarchical $(m_{\nu_1} << 
m_{\nu_2} << m_{\nu_3})$, then $\langle m_{\nu_e} \rangle$ is already 
constrained to be unobservably small \cite{7} from the present data on 
neutrino oscillations.  In this note, I will consider instead the case of 
$\nu_e$ almost degenerate in mass with a linear combination of $\nu_\mu$ and 
$\nu_\tau$, and obtain the interesting prediction that
\begin{equation}
0.20 ~{\rm eV} < m_{\nu_e} \simeq \langle m_{\nu_e} \rangle < 0.36 ~{\rm eV}.
\end{equation}

The idea of three nearly mass-degenerate neutrinos has received a lot of 
attention in the recent literature \cite{8} and is being pursued actively at 
present \cite{9}.  The first task is to identify the mass-generating 
mechanism which is often assumed to be seesaw \cite{10}.  In that case, 
without any further input, the mass splitting of the three neutrinos is due 
to two-loop double $W$ exchange \cite{11} and is suitable \cite{12} for 
vacuum solar neutrino oscillations.  On the other hand, if the alternative 
mass-generating mechanism \cite{13} of a heavy scalar triplet \cite{14} is 
used, the mass splitting occurs in one loop, and a particular choice of 
charged-lepton eigenstates results in a simple formula \cite{15} relating 
atmospheric and vacuum solar neutrino oscillations.  Both of the above two 
specific scenarios \cite{12,15} have in common the assumption that $\langle 
m_{\nu_e} \rangle$ is negligibly small.  Hence they would not be consistent 
with any experimental evidence of a nonzero $\langle m_{\nu_e} \rangle$ in 
the near future.  However, a different choice of charged-lepton eigenstates 
is possible and would correspond to the small-angle matter-enhanced 
oscillation solution \cite{16} of the solar neutrino deficit, resulting in 
Eq.~(2) for $\Delta m^2 \simeq (3-10) \times 10^{-6}$ eV$^2$ \cite {17}, 
as shown below.

The minimal standard model (without right-handed singlet neutrinos) is simply 
extended to include a heavy scalar triplet $\xi = (\xi^{++}, \xi^+, \xi^0)$ 
with $m_\xi^2 >> m_W^2$, to provide the three neutrinos $\nu_e, \nu_\mu, 
\nu_\tau$ with small Majorana masses \cite{14}.  Assume the existence of an 
$SO(2)$ symmetry such that two of the lepton doublets have equal Yukawa 
couplings to $\xi$.  The relevant terms in the interaction Lagrangian are 
then given by
\begin{equation}
{\cal L}_{int} = -\xi^0 [f_0 (\nu_1 \nu_1 + \nu_2 \nu_2) + f_3 \nu_3 \nu_3] 
- \mu \bar \xi^0 \phi^0 \phi^0 + h.c.,
\end{equation}
where $\Phi = (\phi^+,\phi^0)$ is the usual Higgs doublet of the standard 
model. The field $\xi^0$ acquires a naturally small vacuum expectation value 
\cite{14} $u \simeq \mu \langle \phi^0 \rangle^2 / m_\xi^2$ and the 
$3 \times 3$ Majorana neutrino mass matrix is of the form
\begin{equation}
{\cal M}_\nu = \left( \begin{array}{c@{\quad}c@{\quad}c} m_0 & 0 & 0 \\ 
0 & m_0 & 0 \\ 0 & 0 & m_3 \end{array} \right),
\end{equation}
where $m_0 = 2 f_0 u$ and $m_3 = 2 f_3 u$.

The neutrinos must now be identified with their charged-lepton partners. 
The crucial assumption here is that $\nu_1$ is mostly, but not entirely 
$\nu_e$.  Consider first
\begin{equation}
\nu_1 = \nu_e, ~~~ \nu_2 = c \nu_\tau - s \nu_\mu, ~~~ \nu_3 = c \nu_\mu + 
s \nu_\tau,
\end{equation}
where $s \equiv \sin \theta$ and $c \equiv \cos \theta$.  This construction 
is made to accommodate the atmospheric data \cite{1} as $\nu_\mu - \nu_\tau$ 
oscillations with $\sin^2 2 \theta = 4 s^2 c^2$ and $\Delta m^2 = m_0^2 - 
m_3^2$.  To explain the solar data, the two-fold degeneracy of the $\nu_1 - 
\nu_2$ sector is seen to be broken radiatively in one loop.  There are two 
effects. One is a finite correction to the mass matrix, as shown in Figure 1. 
The other is a renormalization of the coupling matrix \cite{19} from the shift 
in mass scale from $m_\xi$ to $m_W$.  As expected, the dominant contribution 
comes from the $\tau$ Yukawa coupling.  The two contributions are naturally 
of the same texture and are easily calculated to be $4I/3$ and $-I/3$ 
respectively, where
\begin{equation}
I = {3 G_F m_\tau^2 \over 16 \pi^2 \sqrt 2} \ln {m_\xi^2 \over m_W^2}.
\end{equation}
The mass matrix ${\cal M}_\nu$ is now corrected to read
\begin{equation}
{\cal M}_\nu = \left( \begin{array}{c@{\quad}c@{\quad}c} m_0 & 0 & 0 \\ 0 & 
m_0 (1+2c^2I) & -sc(m_0+m_3) I \\ 0 & -sc(m_0+m_3) I & m_3 (1+2s^2I) 
\end{array} \right).
\end{equation}
Hence
\begin{equation}
{(\Delta m^2)_{12} \over m_0^2} = 4 c^2 I = {3 c^2 G_F m_\tau^2 \over 4 \pi^2 
\sqrt 2} \ln {m_\xi^2 \over m_W^2}.
\end{equation}
Consider next a small correction to Eq.~(5), {\it i.e.}
\begin{eqnarray}
\nu_1 &=& \nu_e \cos \theta' - (c \nu_\tau - s \nu_\mu) \sin \theta', \\ 
\nu_2 &=& \nu_e \sin \theta' + (c \nu_\tau - s \nu_\mu) \cos \theta',
\end{eqnarray}
then the solar neutrino data may be explained \cite{17} with 
$\sin^2 2 \theta' \simeq (2-10) \times 10^{-3}$ and $(\Delta m^2)_{12} \simeq 
(3-10) \times 10^{-6}$ eV$^2$.  Note that $\nu_2$ is indeed heavier than 
$\nu_1$, as is required for matter enhancement \cite{16} of neutrino 
oscillations in the sun.  Using $c^2 = 0.7$ so that $\sin^2 2 \theta 
= 0.84$ for atmospheric neutrino oscillations and $m_\xi = 10^{14}$ GeV, 
Eq.~(8) then yields Eq.~(2).  Note that a lower value of $m_\xi$ or $c^2$ 
would increase $m_{\nu_e} (\simeq m_0)$, making it even worse for a potential 
conflict with experiment \cite{new}.

In conclusion, I have presented in this note a simple model of neutrino 
masses, using a heavy scalar triplet instead of the canonical seesaw 
mechanism. Assuming equal Majorana masses for $\nu_e$ and a linear 
combination of $\nu_\mu$ and $\nu_\tau$, I have shown that their radiative 
splitting is given by Eq.~(8).  Using the allowed range of 
$(\Delta m^2)_{12}$ values for the small-angle matter-enhanced oscillation 
solution of the solar neutrino deficit, I obtain 0.20 eV $< m_{\nu_e} \simeq 
\langle m_{\nu_e} \rangle <$ 0.36 eV.  Given the most recently reported 
upper limit of 0.2 eV (where the experimental sensitivity is 0.38 eV 
at 90\% C.L.), this prediction is tantalizingly close to being ruled out 
(or confirmed), and is certainly within reach of future experiments.
\vspace{0.3in}
\begin{center} {ACKNOWLEDGEMENT}
\end{center}

This work was supported in part by the U.~S.~Department of Energy under 
Grant No.~DE-FG03-94ER40837.

\bibliographystyle {unsrt}

\begin{thebibliography} {99}
\bibitem{1} Y. Fukuda {\it et al.}, Phys. Lett. {\bf B433}, 9 (1998); 
{\bf B436}, 33 (1998); Phys. Rev. Lett. {\bf 81}, 1562 (1998); {\bf 82}, 2644 
(1999).
\bibitem{2} R. Davis, Prog. Part. Nucl. Phys. {\bf 32}, 13 (1994);  P. 
Anselmann {\it et al.}, Phys. Lett. {\bf B357}, 237 (1995); {\bf B361}, 235 
(1996); J. N. Abdurashitov {\it et al.}, Phys. Lett. {\bf B328}, 234 (1994); 
Y. Fukuda {\it et al.}, Phys. Rev. Lett. {\bf 77}, 1683 (1996); {\bf 81}, 
1158 (1998); {\bf 82}, 1810 (1999); {\bf 82}, 2430 (1999).
\bibitem{3} C. Athanassopoulos {\it et al.}, Phys. Rev. Lett. {\bf 75}, 2650 
(1995); {\bf 77}, 3082 (1996); {\bf 81}, 1774 (1998).
\bibitem{4} A. I. Belesev {\it et al.}, Phys. Lett. {\bf B350}, 263 (1995).
\bibitem{5} L. Baudis {\it et al.}, Phys. Lett. {\bf B407}, 219 (1997).
\bibitem{new} L. Baudis {\it et al.}, hep-ex/9902014.
\bibitem{6} H. V. Klapdor-Kleingrothaus, hep-ex/9901021.
\bibitem{7} S. M. Bilenky, C. Giunti, C. W. Kim, and M. Monteno, Phys. Rev. 
{\bf D57}, 6981 (1998).
\bibitem{8} D. Caldwell and R. N. Mohapatra, Phys. Rev. {\bf D48}, 3259 
(1993);  A. S. Joshipura, Z. Phys. {\bf C64}, 31 (1994); Phys. Rev. {\bf D51}, 
1321 (1995); P. Bamert and C. P. Burgess, Phys. Lett. {\bf B329}, 289 (1994); 
D.-G. Lee and R. N. Mohapatra, Phys. Lett. {\bf B329}, 463 (1994); A. 
Ioannisian and J. W. F. Valle, Phys. Lett. {\bf B332}, 93 (1994); A. Ghosal, 
Phys. Lett. {\bf B398}, 315 (1997); A. K. Ray and S. Sarkar, Phys. Rev. 
{\bf D58}, 055010 (1998); C. D. Carone and M. Sher, Phys. Lett. {\bf B420}, 
83 (1998); H. Fritzsch and Z. Xing, Phys. Lett. {\bf B440}, 313 (1998); U. 
Sarkar, Phys. Rev. {\bf D59}, 037302 (1999); G. C. Branco, M. N. Rebelo, 
and J. I. Silva-Marcos, Phys. Rev. Lett. {\bf 82}, 683 (1999).
\bibitem{9} F. Vissani, hep-ph/9708483; H. Georgi and S. L. Glashow, 
hep-ph/9808293; R. N. Mohapatra and S. Nussinov, hep-ph/9809415; Y. L. Wu, 
hep-ph/9810491, 9901245, 9901320; C. Wetterich, Phys. Lett. {\bf B451}, 397 
(1999); R. Barbieri, L. J. Hall, G. L. Kane, and G. G. Ross, hep-ph/9901228.
\bibitem{10} M. Gell-Mann, P. Ramond, and R. Slansky, in {\it Supergravity}, 
edited by P. van Nieuwenhuizen and D. Z. Freedman (North-Holland, Amsterdam, 
1979), p.~315; T. Yanagida, in {\it Proceedings of the Workshop on the 
Unified Theory and the Baryon Number in the Universe}, edited by O. Sawada 
and A. Sugamoto (KEK Report No.~79-18, Tsukuba, Japan, 1979), p.~95; R. N. 
Mohapatra and G. Senjanovic, Phys. Rev. Lett. {\bf 44}, 912 (1980).
\bibitem{11} K. S. Babu and E. Ma, Phys. Rev. Lett. {\bf 61}, 674 (1988); 
Phys. Lett. {\bf B228}, 508 (1989).  See also S. T. Petcov and S. T. Toshev, 
Phys. Lett. {\bf B143}, 175 (1984).
\bibitem{12} E. Ma, hep-ph/9812344 (Phys. Lett. {\bf B}, in press).
\bibitem{13} E. Ma, Phys. Rev. Lett. {\bf 81}, 1171 (1998).
\bibitem{14} E. Ma and U. Sarkar, Phys. Rev. Lett. {\bf 80}, 5716 (1998). 
\bibitem{15} E. Ma, hep-ph/9902392.
\bibitem{16} L. Wolfenstein, Phys. Rev. {\bf D17}, 2369 (1978); S. P. 
Mikheyev and A. Yu. Smirnov, Sov. J. Nucl. Phys. {\bf 42}, 913 (1986).
\bibitem{17} J. N. Bahcall, P. I. Krastev, and A. Yu. Smirnov, Phys. Rev. 
{\bf D58}, 096016 (1998).
\bibitem{19} J. Ellis and S. Lola, hep-ph/9904279; J. A. Casas, J. R. 
Espinosa, A. Ibarra, and I. Navarro, hep-ph/9904395.
\end{thebibliography}

\begin{center}
\begin{picture}(360,200)(0,0)
\ArrowLine(30,0)(90,0)
\Text(60,-8)[c]{$\nu_i$}
\ArrowLine(180,0)(90,0)
\Text(135,-8)[c]{$\tau_L$}
\ArrowLine(270,0)(180,0)
\Text(225,-8)[c]{$\tau_R$}
\ArrowLine(330,0)(270,0)
\Text(300,-8)[c]{$\nu_\tau$}
\DashArrowLine(180,-40)(180,0)6
\Text(180,-50)[c]{$\langle \phi^0 \rangle$}
\DashArrowLine(180,97)(180,57)6
\Text(180,106)[c]{$\langle \phi^0 \rangle$}
\DashArrowArcn(180,-45)(101,154,90)6
\Text(130,52)[c]{$\xi^-$}
\DashArrowArcn(180,-45)(101,90,26)6
\Text(240,52)[c]{$\phi^-$}
\end{picture}
\vskip 2.0in
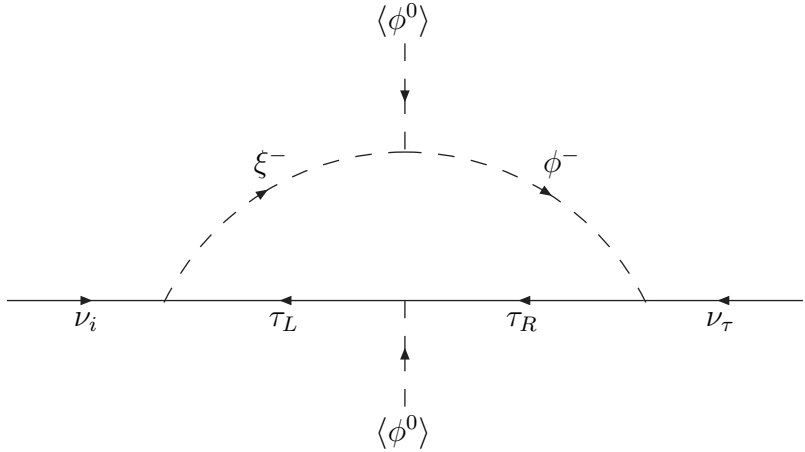
{\bf Fig.~1.} ~ One-loop radiative breaking of neutrino mass degeneracy.
\end{center}
\end{document}